# Merging the CEM2k and LAQGSM Codes with GEMINI

M. I. Baznat*, S. G. Mashnik†, K. K. Gudima,* and R. E. Prael†

*Institute of Applied Physics, Academy of Science of Moldova, Chişinău, MD-2028, Moldova
†Los Alamos National Laboratory, Los Alamos, NM 87545, USA

**Abstract.** An improved version of the Cascade-Exciton Model (CEM) of nuclear reactions contained in the code CEM2k and the Los Alamos version of the Quark-Gluon String Model (LAQGSM) are merged with the well-known sequential-binary-decay model GEMINI by Charity. We present some results on proton-induced fragmentation, fission-product yields and on particle spectra predicted by these extended versions of CEM2k and LAQGSM. We show that merging CEM2k and LAQGSM with GEMINI allows us to describe many fission and fragmentation reactions in addition to the spallation and evaporation reactions which are already described well by these codes. Nevertheless, the current version of GEMINI we use does not provide a completely satisfactory description of some complex-particle spectra, fragment emission, and spallation yields for some reactions, and is not yet a universal tool for applications. Our results show that GEMINI contains a powerful model to describe evaporation/fission/fragmentation reactions and often provides better results when compared to other models, especially for emission of heavy fragments from reactions on medium-heavy nuclei (where most other models simply fail), but it must be further extended and improved in order to properly describe arbitrary reactions.

## INTRODUCTION

Recently, we have developed at Los Alamos National Laboratory an improved version of the Cascade-Exciton Model (CEM) of nuclear reactions contained in the codes CEM2k [1] and the Los Alamos version of the Quark-Gluon String Model (LAQGSM) [2] to describe reactions induced by particles and nuclei at energies up to hundreds of GeV/nucleon for a number of applications. Originally, both CEM2k and LAQGSM were not able to describe fission and production of light fragments heavier than $^4$He, as they had neither a fission nor a fragmentation model. We have earlier addressed this problem by improving these codes and merging them with several evaporation/fission/fragmentation models. One of the more promising approaches from the point of view of describing many available measurements involved merging [3–5] CEM2k and LAQGSM with the Generalized Evaporation/fission Model code GEM2 by Furihata [6]. CEM2k+GEM2 and LAQGSM+GEM2 perform as well as and often better than other current models in describing a large variety of spallation, fission, and fragmentation reactions (see, *e. g.*, [7, 8] and references therein). Neveretheless, these versions of the codes fail to reproduce correctly production of fission-like heavy fragments from reactions with medium and light nuclei (see Fig. 1). Such nuclear targets are considered too light to fission in conventional codes (including GEM2 and all models currently employed in large-scale transport codes). Similarly, the fragments are too light to be produced as spallation residues and too heavy to be produced via standard evaporation models.

One way to approach this problem is to describe the fast part of a nuclear reaction with an IntraNuclear Cascade model (INC) followed by preequilibrium emission of particles during the equilibration of the excited residual nucleus. At this point, one would employ a fission-like sequental-binary-decay model, like the well-known code GEMINI by Charity [9], to describe the compound-nuclear decay. In our case, this means separately merging CEM2k and LAQGSM with GEMINI. We have done this and present some illustrative results from the merged codes.

## RESULTS AND DISCUSSION

The CEM2k and LAQGSM models are described in detail in [1, 2] and references therein. The GEMINI code is available for downloading from the Web and is described in detail in [9]; therefore we only outline the main ideas of these models. CEM2k and LAQGSM describe the first, fast part of a reaction in terms of a space-like or a time-like intranuclear cascade model, respectively. The excited residual nucleus remaining after the cascade determines the particle-hole configuration that is the starting point for the preequilibrium stage of the reaction. The subsequent relaxation of the nuclear excitation is treated in terms of an improved Modified Exciton Model (MEM) of preequilibrium decay followed by the equilibrium evaporative final stage of the reaction. We use GEMINI to describe the evaporation/fission/fragmentation of excited nuclei produced after the preequilibrium stage.

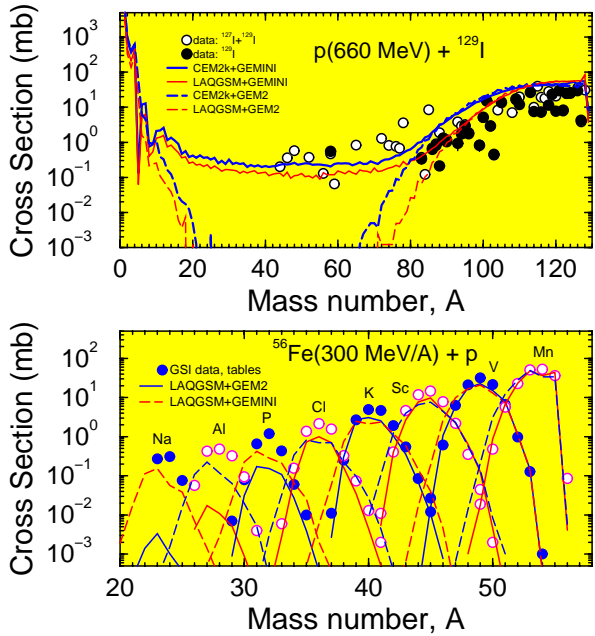
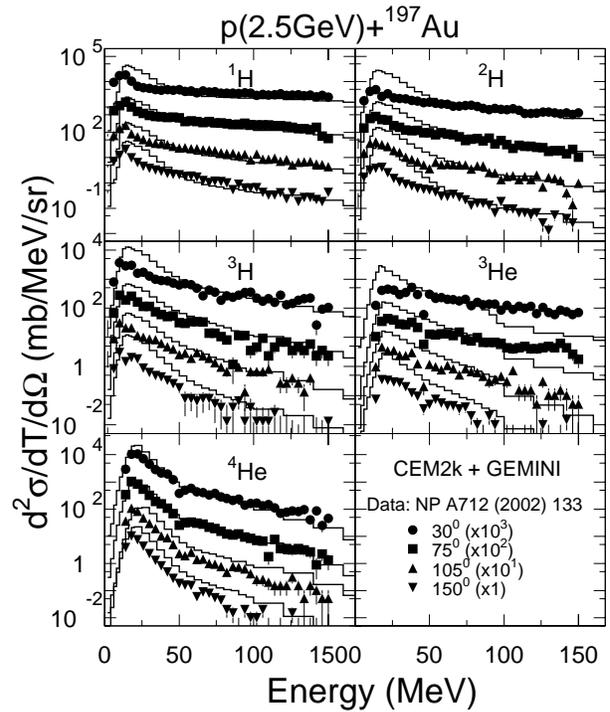

**FIGURE 1.** Top plot: Experimental data on 660 MeV p + $^{129}$I (and $^{127}$I + $^{129}$I) [12] compared with mass distributions of products predicted by CEM2k and LAQGSM merged with GEM2 (dashed lines) and GEMINI (solid lines), as indicated. Bottom plot: Experimental [13] mass distributions of the yields of eight isotopes from Mn to Na produced in the reaction 300 MeV/A $^{56}$Fe + p compared with LAQGSM+GEM2 (solid lines) and LAQGSM+GEMINI (dashed lines) results. *t_delay* = 0.1 and *sig_delay* = 0.1 are used in GEMINI to calculate both these reactions.

**FIGURE 2.** Experimental spectra of p, d, t, $^3$He and $^4$He emitted from the reaction 2.5 GeV p + $^{197}$Au at 30, 75, 105, and 150 degrees [16] compared with results by CEM2k+GEMINI, as indicated. *t_delay* = 17 and *sig_delay* = 10 are used in GEMINI to calculate this reaction.

The code GEMINI [9] calculates decay of the compound nucleus by subsequent binary fission-like decays. All possible decay modes, from emission of nucleons and light fragments through asymmetric to symmetric fission are considered. The decay width for the evaporation of fragments with $Z \leq 2$ is calculated using the Hauser-Feshbach formalism [10]. For binary divisions corresponding to the emission of heavier fragments, the decay width is calculated using the transition state formalism of Moretto [11]. We use here the default version of GEMINI, without any changes or fitting of parameters, except the value of the delay time for fission (GEMINI input parameters *t_delay* and *sig_delay*; the values of both these parameters used here are listed in figure captions). The level density parameter is taken fixed as $a = A/12$ MeV$^{-1}$ for both the residual nucleus and the saddle-point transition state. Further details on GEMINI may be found in [9] and references therein.

Fig. 1 shows two examples of intermediate-energy proton-induced reactions on medium-mass nuclei. We see that CEM2k and LAQGSM merged with GEM2 fail to reproduce the yield of fission-like heavy fragments, while CEM2k+GEMINI and LAQGSM+GEMINI describe these fragments reasonably well, considering no fitting or adjustments of any parameters are done for these calculations.

The reactions shown in Fig. 1 (and other similar reactions) can be described also by versions of CEM2k and LAQGSM when they are merged [14] with the statistical multifragmentation model code SMM by Botvina *et al.* [15] instead of GEMINI. This makes it more difficult to determine the mechanism of such processes. We think that for such intermediate-energy proton-induced reactions the contribution of multifragmentation to the production of heavy fragments should not be not very significant due to the relatively low excitation energies involved. Such fragments are more likely to be produced via the fission-like binary decays modeled by GEMINI.

The spectra of emitted particles are not too well described by CEM2k+GEMINI and LAQGSM+GEMINI but are still reasonable. One example is shown in Figs. 2 and 3, for the reaction 2.5 GeV *p* + Au. We see that CEM2k+GEMINI overestimates significantly the low-energy part of the measured spectra of *p*, *d*, *t* and $^3$He, while not so much the $^4$He spectra. We believe that this is not critical; we expect that slightly adjusting the parameters of CEM2k+GEMINI and LAQGSM+GEMINI could lead to a reasonable result, but this is not an aim of this work. The overestimation of the high-energy tails of the $t$, $^3$He, and $^4$He spectra at backward angles is due mainly to preequilibrium emission, as one can deduce from Fig. 3, and has nothing to do with GEMINI. This problem could be addressed by improving the preequilibrium model, but this is again outside the scope of this work.

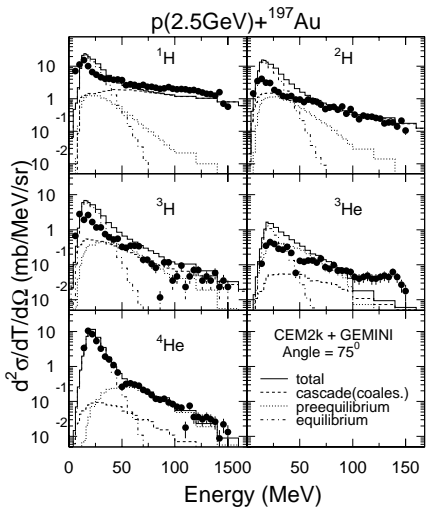

**FIGURE 3.** The spectra at 75 degrees from Fig. 2, with solid histograms showing CEM2k+GEMINI total results, while dashed, dotted, and dot-dashed histograms show the contributions to the total from INC (coalescence, in the case of complex particles), preequilibrium, and equilibrium (from GEMINI) processes, respectively. $t\_delay = 17$ and $sig\_delay = 10$ are used in GEMINI to calculate this reaction.

Fig. 4 shows an example of proton-induced nuclide production on preactinides, namely 1 GeV $p$ + $^{208}$Pb, measured at GSI in inverse kinematics [17]. We see that LAQGSM+GEMINI (as does also CEM2k+GEMINI) describes very well the fission-fragment yields, and also the yields of many of the spallation products. But the farther from the target we go, the poorer the agreement of calculations with the data; in the region between spallation and fission products (from Tb to In) LAQGSM+GEMINI does not describe correctly the measured cross sections. This is more striking in Fig. 5, where we show the mass distribution of all products from the same reaction. Similar results are obtained for other reactions.

The problem increases when we move to heavier actinide targets, where CEM2k+GEMINI and LAQGSM+GEMINI do not describe properly both the spallation and the fission products. This is because GEMINI was developed and should only be applied to describe not-too-heavy nuclei: Fig. 6 demonstrates that GEMINI completely fails to describe the double-humped mass distribution of fission fragments from actinides that one finds experimentally.

Our results show that merging CEM2k and LAQGSM with GEMINI allows us to describe many fission and fragmentation reactions, especially on targets below the actinide region. We found that: 1) GEMINI merged with CEM2k/LAQGSM provides reasonably good results for medium-heavy targets without a fission delay time; 2) For preactinedes, we have to use $t\_delay = 50$–70 and $sig\_delay = 1$–50, otherwise GEMINI provides too much fission—this may be related to the calculation of fission barriers of preactinides with strong ground-state shell corrections in the version of GEMINI we used; 3) The current version of GEMINI does not work well for actinides (see Fig. 6).

The current version of GEMINI merged with CEM2k and LAQGSM also does not provide a completely satisfactory description of some complex-particle spectra, fragment emission, and spallation yields for some reactions, and is not yet a universal tool for applications. GEMINI contains a powerful approach to describe evaporation/fission/fragmentation reactions and often provides better results in comparison with other modern models, especially for emission of heavy fragments from reactions on medium-heavy nuclei (where most other models simply fail), but it must be further extended and improved in order to properly describe arbitrary reactions.

## ACKNOWLEDGEMENTS

We thank Dr. Arnold Sierk for useful discussions, help, and liaison with Prof. Robert Charity, the author of GEMINI. We are grateful to Dr. Claude Volant for noticing that the value of the delay time for fission in GEMINI was given in error in the intitial version of this paper for several reactions. This work was partially supported by the US Department of Energy, Moldovan-US Bilateral Grants Program, CRDF Project MP2-3045, and the NASA ATP01 Grant NRA-01-01-ATP-066.

## REFERENCES

1. Mashnik, S. G. and Sierk, A. J., in *Proc. AccApp00, Washington, DC, USA, 2000*, ANS, La Grange Park, IL, 2001, pp. 328–341; E-print: nucl-th/0011064.
2. Gudima, K. K., Mashnik, S. G., and Sierk, A. J., LANL Report LA-UR-01-6804, Los Alamos (2001).
3. Mashnik, S. G., Gudima, K. K., and Sierk, A. J., in *Proc. SATIF-6, SLAC, USA, 2002*, NEA/OECD, Paris, 2004, pp. 337–366; E-print: nucl-th/0304012.

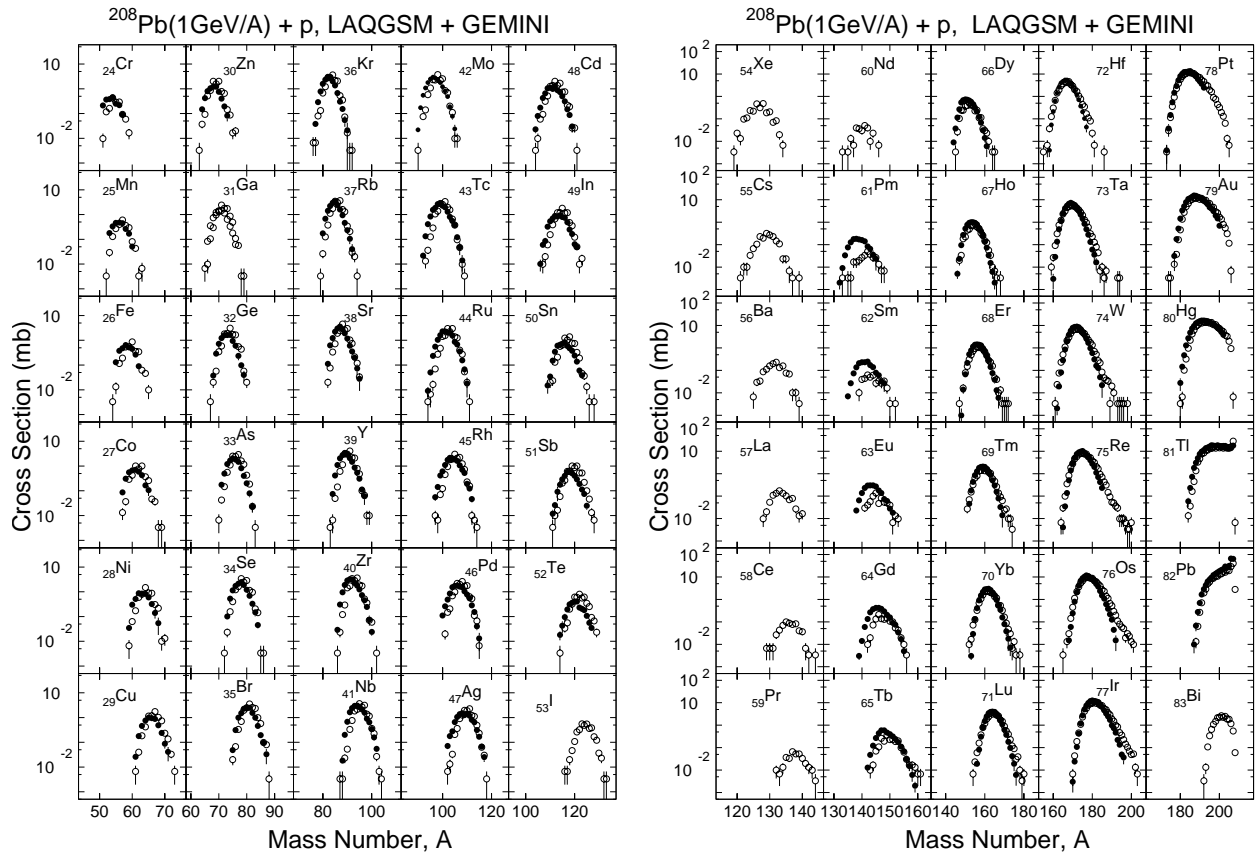

**FIGURE 4.** Comparison of all measured [17] cross sections of fission (left plot) and spallation (right plot) products from the reaction 1 GeV/nucleon $^{208}$Pb on $p$ (filled circles) with LAQGSM+GEMINI results (open circles). Experimental data for products from Nd to I are not available, so we present only our predictions. *t_delay* = 75 and *sig_delay* = 50 are used in GEMINI to calculate this reaction.

4. Mashnik, S. G., Sierk, A. J., and Gudima, K. K., LANL Report LA-UR-02-5185, Los Alamos (2002); E-print: nucl-th/020804.
5. Baznat, M. I., Gudima, K. K., and Mashnik, S. G., in *Proc. AccApp'03, San Diego, USA, 2003*, ANS, La Grange Park, IL, 2004, pp. 976–985; E-print: nucl-th/0307014.
6. Furihata, S., Ph.D. thesis, Tohoku University, Japan (2003); *Nucl. Instr. Meth. B* **171**, 252–258 (2000).
7. Mashnik, S. G., Gudima, K. K., Prael, R. E., and Sierk, A. J., in *Proc. TRAMU@GSI, Darmstad, Germany, 2003*, Eds. A. Kelic and K.-H. Schmidt, ISBN 3-00-012276-1, http://ww-wnt.gsi.de/tramu; E-print: nucl-th/0404018.
8. Mashnik, S. G., Gudima, K. K., Moskalenko, I. V., Prael, R. E., and Sierk, A. J., *Advances in Space Research* **34**, 1288–1296 (2004); E-print: nucl-th/0210065.
9. Charity, R. J. et al., *Nucl. Phys. A* **483**, 371–405 (1988); *Phys. Rev. C* **63**, 024611 (2001) and private communication from R. J. Charity to A. J. Sierk, 2000; http://wunmr.wustl.edu/ rc/.
10. Hauser, H. and Feshbach, H., *Phys. Rev.* **87**, 366–373 (1952).
11. Moretto, L. G., *Nucl. Phys. A* **247**, 211–230 (1975).
12. Adam, J. et al., *Pis'ma v EChAYa [Particles and Nuclei, Letters]* **1**, #4 [121], 53–64 (2004); E-print: nucl-ex/0403056; in *Proc. ND2004, Santa Fe, USA, 2004*, Eds. R. C. Haight, M. B. Chadwick, T. Kawano, and P. Talou, Melville: AIP Conf. Procs. vol 769, 2005, pp. 1047–1050.
13. Villagrasa-Canton, C., Ph.D. thesis, Universite de Paris XI Orsay, (2003); in *Proc. ND2004, Santa Fe, USA, 2004*, Eds. R. C. Haight, M. B. Chadwick, T. Kawano, and P. Talou, Melville: AIP Conf. Procs. vol 769, 2005, pp. 842–845; www-w2k.gsi.de/charms/Preprints/ND-2004-Santa-Fe/villagrasac.pdf.
14. Mashnik, S. G., Gudima, K. K., Sierk, A. J., and Prael, R. E., LANL Research Note X-5-RN(U)04-08 (2004); in *Proc. ND2004, Santa Fe, USA, 2004*, Eds. R. C. Haight, M. B. Chadwick, T. Kawano, and P. Talou, Melville: AIP Conf. Procs. vol 769, 2005, pp. 1188–1192; E-print: nucl-th/0502019.
15. Bondorf, J. P., Botvina, A. S., Iljinov, A. S., Mishustin, I. N., and Sneppen, K., *Phys. Rep.* **257**, 133–221 (1995).
16. Letourneau, A. et al., *Nucl. Phys. A* **712**, 133–166 (2002).
17. Enqvist, T. et al., *Nucl. Phys. A* **686**, 481–524 (2001).

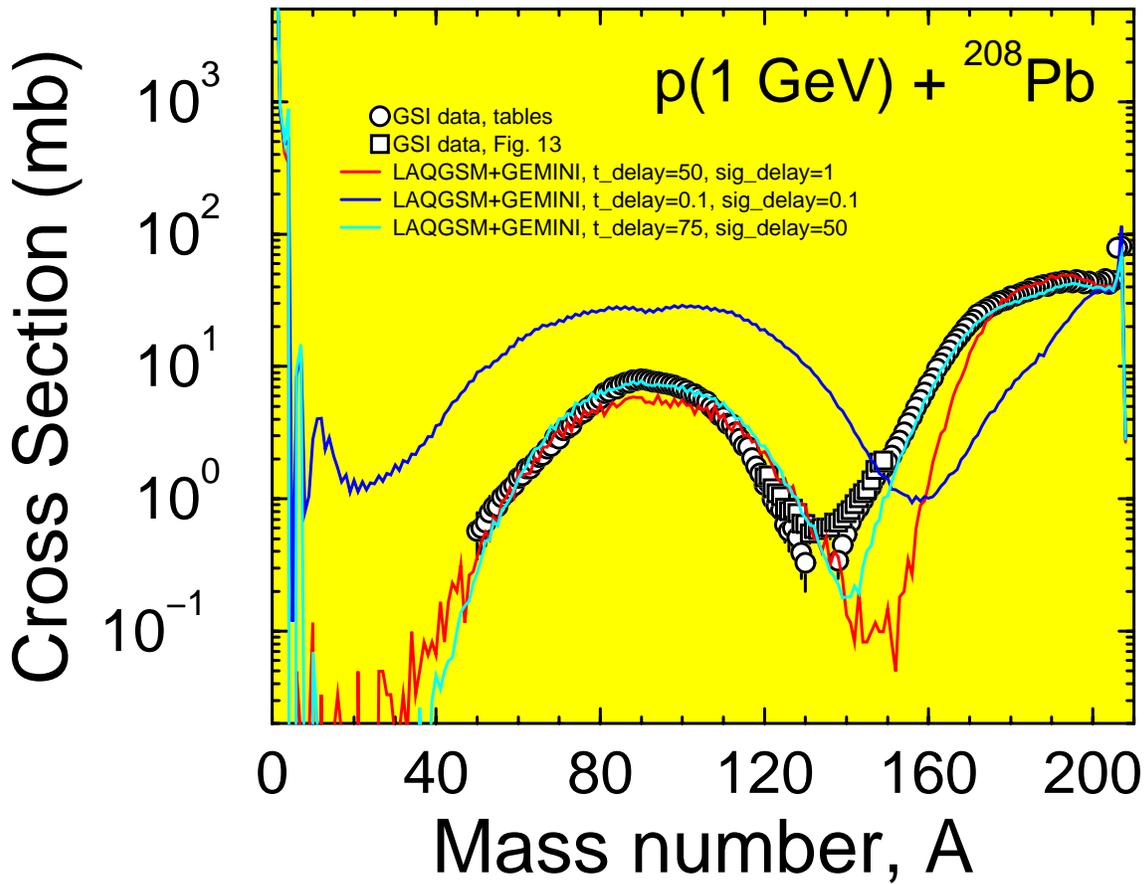

**FIGURE 5.** Summed mass distribution of all elements for the same reaction and the same experimental data as shown in Fig. 4. Results obtained with three different values of GEMINI input parameters *t_delay* and *sig_delay* indicated in legend are shown here.

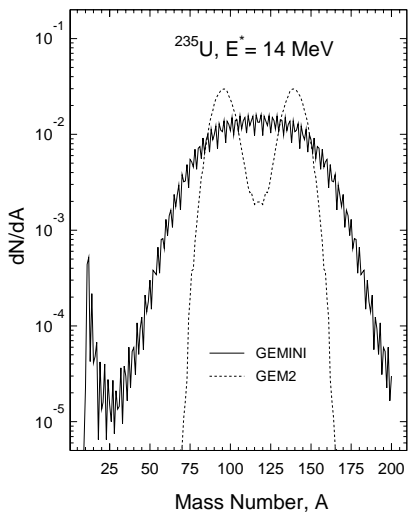

**FIGURE 6.** Mass distributions of fission fragments from the compound nucleus $^{235}$U with an excitation of 14 MeV as predicted by GEMINI (solid line) and GEM2 (dashed line), respectively. No delay time for fission is used here.